\documentclass[12pt,english]{article}
\usepackage{babel}
\usepackage{amsfonts}
\title{An effective interaction spontaneously \\
arising in a renormalizable model\\
of quantum field theory}
\author{B.A. Arbuzov\\
{\it Skobeltsyn Institute for Nuclear Physics 
of MSU}\\ {\it 119992 Moscow, RF}\\
E-mail: arbuzov@theory.sinp.msu.ru}
\date{}
\newcommand{\be}{\begin{equation}}
\newcommand{\ee}{\end{equation}}
\newcommand{\beq}{\begin{eqnarray}}
\newcommand{\eeq}{\end{eqnarray}}       
\newcommand{\nn}{\nonumber}
\newcommand{\bi}{\bibitem}
\begin{document}
\maketitle
\begin{quote}
Theory of massless scalar field $\phi$ 
with interaction $g \phi^3$  
in six-dimensional space is considered.
A possibility of initial scale invariance breaking, which  
results in a spontaneous arising of effective interaction 
$G\,\phi^4$, is studied by application of 
Bogolubov quasi-averages approach.  
It is shown, that compensation equation for 
form-factor of this interaction in approximation 
up to the third order in $G$ has a non-trivial 
solution. In the same approximation Bethe-Salpeter equation 
for a zero-mass bound state of two scalar fields $\phi$ is 
shown to have a solution. The conditions imposed on form-factor 
value at 
zero and scalar field mass $m$ fix the unique solution, 
which gives relations between parameters of interaction $g \phi^3$ 
and parameters $G\,$ and $m$. Arguments are laid down 
in favour of a stability of the non-trivial solution.
\end{quote}


\section{Compensation equation in quasi-averages 
approach}

N.N. Bogolubov quasi-averages method~\cite{Bog2, Bog} is the most
consistent and effective method of studying of a spontaneous 
symmetry breaking problems.  

An important point of the quasi-averages method is connected 
with a compensation equation~\cite{Bog2, Bog}. Bearing in 
mind applications in the present work let us briefly 
formulate method of construction of the compensation 
equations. In the line of a study of a possible 
spontaneous symmetry breaking in quantum field theory 
problems in method~\cite{Bog} the following procedure is 
applied\footnote{At first methods~\cite{Bog2, Bog}
where applied to quantum theory problems in 
work~\cite{ATF}}. Let the initial Lagrangian  
\be
L\,=\,L_0\,+\,L_{int}\,;\label{0+int}
\ee
to possess some symmetry. Let us add to expression~(\ref
{0+int}) some term $\epsilon\,L_{br}$, which breaks the 
initial symmetry. With this modification of the problem 
we perform evaluations of necessary quantities and 
we set $\epsilon \to 0$ only after these 
evaluations. Not always the results of such a procedure 
(quasi-averages) coincide with results, obtained 
in the framework of the initial symmetric problem 
(simply averages). In the line of these 
evaluations of quasi-averages one has to solve compensation 
equations. For instance, in a theory with the initial chiral 
symmetry fermions are to have zero masses. Let us use the 
following small increment which breaks the symmetry 
\be
\epsilon\,L_{br}\,=\,-\,\epsilon\,\bar \psi\,\psi\,.
\label{atf}
\ee
Now let us add to the modified Lagrangian~(\ref{atf}) 
a possible mass term and subtract the same. We have
\be
L\,=\,L_0\,-\,m\,\bar \psi\,\psi\,+\,L_{int}\,+\,m\,
\bar \psi\,\psi\,-\,\epsilon\,\bar \psi\,\psi\,\label{dv}
\ee
Let the fist two terms to be the new free 
Lagrangian while the three last terms now comprise 
the new interaction Lagrangian. Then we have to demand 
the new interaction does not contribute to the mass term, 
that is two-field Green function obtained from the 
modified interaction Lagrangian be zero on the mass 
shell. This condition is just the compensation equation 
of the problem. In the case under consideration this 
condition leads to equation
\be
-\,m\,+\,\epsilon\,+\,\Sigma(m)\,=\,0\,;\label{comp}
\ee
where $\Sigma(m)$ is mass operator on the mass shell of 
the modified free Lagrangian. In this equation one already 
can set $\epsilon \to 0$. As a rule (see e.g.~\cite{ATF}) 
mass operator $\Sigma(m)$ is proportional to $m$ and trivial 
solution of the compensation equation $m\,=\,0$ always 
exists. However nontrivial solutions $m\,\ne\,0$ also 
may exist.

Thus the main principle of construction of a compensation 
equation consist in the procedure "add -- subtract"$\,$
of symmetry breaking terms, 
one of these terms being related to the free 
Lagrangian and the other one being related to the 
interaction Lagrangian. 
Then one has to compensate that term, which is to be zero in 
the corresponding problem. This principle will be applied 
in the present work.

\section{Justification of the model choice}

The phenomenon of spontaneous symmetry breaking is decisive 
for formulation of the electroweak theory. The introduction of 
elementary scalar fields~\cite{Higgs} is well-known to be the essence 
of the standard way of the breaking. Their self-interaction leads to 
redefinition of the vacuum in the theory and to existence of 
the scalar Higgs particle. However proposals based on a dynamical 
breaking of the initial symmetry without elementary Higgs scalars 
are also considered. This leads to effective four-fermion 
interaction of heavy quarks either to be postulated (see e.g. 
review~\cite{t}) or to be dynamically arisen~\cite{Arb1, SUSY}. 
As a result the initial theory, which contains scale-invariant 
gauge interactions, transforms into a theory, which contains 
interactions with dimensional coupling constant as well, that 
explicitly breaks the scale invariance. In works being 
mentioned~\cite{Arb1, SUSY} assumption was made on a possibility 
of existence of solutions of corresponding compensation equations. 
However this assumption was not duly justified. The purpose of the 
present work is to consider a simpler model, which would allow 
to have exact solutions of (approximate) compensation equations. 
Using these solutions one could study conditions under which 
the assumptions would be fulfilled. To some extent the model has to 
correspond to features of the electroweak theory. Namely we 
achieve a simplicity by considering a scalar field. In view of 
coupling constants to have proper dimensions we choose dimensionality 
of the space-time to be six. Really, in this case the coupling 
constant 
of interaction $g\,\phi^3$ is dimensionless and interaction 
$G\,\phi^4$ 
has constant of inverse mass squared dimension, that corresponds to 
the dimension of a constant of a four-fermion interaction in 
four-dimensional space-time.

So we introduce in the six-dimensional space-time a scalar field 
$\phi$ 
with initiative scale-invariant Lagrangian
\be
L\,=\,\frac{1}{2}\,g^{\mu \nu}\,\frac{\partial \phi}{\partial x^\mu}
\frac{\partial \phi}
{\partial x^\nu}\,+\,\frac{g_0}{3!}\,\phi^3\,.\label{init}
\ee

Let us choose the natural signature with one time and five space 
axes. 
The transition from this space-time to Euclidean six-dimensional 
space 
is accompanied by the following substitutions
\be
p^2\,\rightarrow\,-\,p^2_E\,;\qquad
d^6\,p\,\rightarrow\,\imath\,d^6_E\,p\,.\label{eucl}
\ee
It was important for us to find a model, which corresponds 
to the approach under consideration. So in the present work 
we will not discuss physical meaning of a multi-dimensional 
theory and we consider the chosen variant as purely model one, 
as well as two-dimensional models are often considered.

Now we start with Lagrangian~(\ref{init}). Evident evaluations 
give one-loop renormalization group equation~\cite{BSh} 
for $g^2(\mu^2)$
\be
\frac{d\,g^2(\mu^2)}{d L}\,=\,-\,\frac{3\,g^4}{4\,(4\,\pi)^3}\,;
\qquad 
L\,=\,\log\frac{\mu^2}{\Lambda^2_3}\,;\label{GL}
\ee
Solution of equation~(\ref{GL}) has a form
\be
g^2(\mu^2)\,=\,g_0^2\,\biggl(1\,+\,\frac{3\,g_0^2}{4\,(4\,\pi)^3}\,
\log\frac{\mu^2}{\Lambda^2_3}\biggr)^{-1}\,.\label{g2}
\ee
Sometimes it is convenient to use parameter $\bar h(\mu^2)$ defined 
by the following relation
\be
\bar h(\mu^2)\,=\,\frac{3\,g^2(\mu^2)}{4\,(4\,\pi)^3}\,=\,\biggl
(\log\frac{\mu^2}{\Lambda_g^2}\biggr)^{-1}\,;\label{h}
\ee
where for transition from $\Lambda^2_3$ to $\Lambda_g^2$ we 
have used the standard tool analogous to that in QCD:
$$
\Lambda_g^2\,=\,\Lambda_3^2\,\exp \Biggl(-\, \frac{4\,(4 \pi)^3}
{3\,g_0^2} \Biggr)\,.
$$
Thus we get convinced, that the theory~(\ref{init}) is an 
asymptotically free one and expression~(\ref{h}) makes sense 
for $\mu^2\,\gg \,\Lambda_g^2 $. 

Note that in this theory there are quadratic divergences in the 
scalar field mass. It is the common feature of theories with 
elementary scalars. The problem of mass of the scalar field will 
be considered in details later on.

\section{Compensation equation in a six-dimensional scalar 
model}

Let us have a massless scalar field of the six-dimensional 
space. The initial free 
Lagrangian possesses scale symmetry. We shall look for 
a solution, which breaks this symmetry, with the aid of 
Bogolubov quasi-averages approach~\cite{Bog}. Namely 
according to~\cite{Bog} we add to the Lagrangian the 
following small increment
$$
-\,\epsilon\,\frac{\phi^4}{4!}\,.
$$
Now the scale invariance is already broken and an appearance 
of nonlocal terms of the form
\be
G\,\int \bar F(x,x_1,x_2,x_3,x_4)\,\phi(x_1)\phi(x_2)\phi(x_3)
\phi(x_4)\,dx_1\,dx_2\,dx_3\,dx_4;\label{phix}
\ee
is possible. Here $G$ is a dimensional coupling constant and  
$\bar F(x,x_1,x_2,x_3,x_4)$ 
is a function of four differences of coordinates $x-x_i$, which 
Fourier transform 
$F(p_1,p_2,p_3,p_4)$, where $p_i$ are momenta of legs, 
represents a form-factor, defining range of 
interaction~(\ref{phix}). We shall look for a solution, 
decreasing at momentum infinity and thus defining a region 
of action of the effective interaction.

Let us add to the initial free Lagrangian such a term with 
an interaction of the forth power and subtract the same
\be
L\,=\,\frac{1}{2}\,g^{\mu \nu}\,\frac{\partial \phi}{\partial x^\mu}\,
\frac{\partial \phi}{\partial x^\nu}\,-\,\frac{m^2}{2}\,\phi^2\,-\,
\frac{G}{4!}\,F\cdot\phi^4\,\,-\frac{\epsilon}{4!}\phi^4\,
+\,\frac{g_0}{3!}\,\phi^3\,+\,\frac{G}{4!}\,F\cdot\phi^4\,
+\,\frac{m^2}{2}\,\phi^2\,;\label{phi}
\ee
where we use abbreviated notation $-\,G\,F\cdot\phi^4$ 
instead of expression~(\ref{phix}). Of course the presence 
of term~(\ref{phix}) explicitly breaks the scale invariance, 
so we perform a procedure "add -- subtract" for a mass term 
as well. 
Let us refer the 
forth power term with the plus sign to the interaction 
Lagrangian and the same term with the minus 
sign we refer to the free Lagrangian. 
\beq
& & L_0\,=\,\frac{1}{2}\,g^{\mu \nu}\,\frac{\partial \phi}{\partial 
x^\mu}\,
\frac{\partial \phi}{\partial x^\nu}\,-\,\frac{m^2}{2}\,\phi^2\,-\,
\frac{G}{4!}\,F\cdot\,\phi^4\,-\frac{\epsilon}{4!}\phi^4\,;\nn\\
& & L_{int}\,=\,\frac{g_0}{3!}\,\phi^3\,+\,\frac{G}{4!}\,F\cdot
\phi^4\,
+\,\frac{m^2}{2}\,\phi^2\,;\label{0i}
\eeq
According to 
the quasi-averages approach~\cite{Bog} the term with the 
plus sign has to be compensated. This means, that the 
new free Lagrangian leads to zero four-particle connected 
Green functions and as a final result contains only terms 
of the second power in fields. 
Thus performing evaluations with sign which is inherent to 
the term in the new free Lagrangian, we come to the 
compensation equation, which schematically looks in 
the following way: the first order term plus one-loop 
terms plus two-loop terms etc. Emphasize once more, that 
here one has to use term $+\,G\,\phi^4$ as an interaction 
Lagrangian. One has to equalize to zero the expansion 
obtained in such a way . This condition is an equation for 
function $F(p_1,p_2,p_3,p_4)$. We set $\epsilon \to 0$ after 
evaluations, in our case this 
means after compensation equations being obtained.

The equation explicitly 
differs from expansion in powers of interaction Lagrangian
\be
L_{int}\,=\,\frac{G}{4!}\,F\cdot\,\phi^4 \,;\label{int}
\ee
in the sign of the interaction constant. In view of this note let us 
emphasize, that the procedure being described can be applied only to 
symmetry breaking terms of even powers in fields. For terms of odd 
powers, e.g. for three-linear ones, a fulfillment of a 
compensation equation leads to vanishing of connected Green function, 
which is defined by an interaction Lagrangian, because the two 
expansions in this case differ only in overall sign.

Note, that the presence of term $-\,G\,\phi^4$ in the new free 
Lagrangian may lead to appearance of connected Green 
functions of higher powers in $\phi$, that is of the sixth 
power, 
of the eighth power etc. Generally speaking, one has to 
construct a chain set of compensation equations for all these 
Green functions. We start with an equation for 
the fourth power Green function and the problem of higher 
Green functions will be discussed in what follows. 

Let us construct an approximated equation for the fourth 
power connected Green function. First of all we choose the 
following kinematics: both left legs have zero momenta and 
the right ones have momenta $p$ and $- p$. We restrict 
ourselves by terms up to two-loop ones inclusively. 
Namely, we have the first order term -- the point; 
three terms of the second order -- simple loops, i.e. 
a horizontal one and two vertical ones with permuted 
left legs; in the third order we have a horizontal and two 
vertical two-loop chains and six terms "wine glass": 
horizontal wine glasses having bases to the left and to the 
right and vertical ones with bases up and down. The number 
of the last terms is to be counted twice due to permutations 
of the left-sided momenta $p$ and $- p$. Generally speaking, 
in each vertex form-factor $F$ is present. However we can 
solve only a linear version of the equation, which is 
obtained by keeping in the equation the first and the second 
order terms, the two-loop horizontal chain and the wineglass 
with the basing to the right. Contributions of the rest 
third order terms we shall consider later on. 
We proceed to the linear equation  
keeping form-factor $F(p,-p,0,0) \equiv F(p^2)$ in the first 
order term and in right-hand vertices of the horizontal 
loop of the second order, of the horizontal two-loop chain 
and of the wineglass in the third order. Other vertices in diagrams 
we consider to correspond to point-like interaction in which 
the form-factor is changed for its value at zero $(F(0)\,=\,1)$
\be
\frac{G}{4!}\,F(0)\,\phi^4\,=\,\frac{G}{4!}\,\phi^4\,.\label{point}
\ee
In vertical simple 
loops, which as well serve as a kernel of the integral 
equation, we substitute point-like vertices~(\ref{point}). 
Corresponding integrals diverge of course. In view of our 
search for decreasing solutions at momentum infinity  for $F(p^2)$, 
we introduce some cut-off $\Lambda$, which 
existence is to be confirmed by results of a solution of the 
equation. In doing this we make the following substitution
$$
\int_0^\infty dq^2\,\rightarrow\,\Lambda^2\,.
$$
For estimation of $\Lambda$ order of magnitude we use 
the following definition
\be
\Lambda^2\,=\,\int_0^\infty\,F(y)\,dy\,;\label{def}
\ee
where one of vertices is changed for the form-factor. 
For justification of the approach the problem of convergence 
of the integral in~(\ref{def}). We shall use the same cut-off 
$\Lambda$ 
in logarithmically diverging integrals. A possible difference of 
an actual cut-off in these integrals from $\Lambda$ leads to 
some change in constant term $c$< which enters into corresponding 
expressions. It will come clear, that the solution will not depend 
on a value of this constant. Thus the formulation of the equation 
in the framework of the accepted approximations does not contain 
arbitrary assumptions.

We consider the equation in six-dimensional Euclidean space 
with the aid of substitutions~(\ref{eucl}). 
In the course of evaluations one has to perform angle 
integrations in six-dimensional space of functions 
$((p-q)^2)^{-1}$ and $\log (p-q)^2$ with powers of  
$(pq)$. We have (for the logarithmic case see~\cite{Mar})
\beq
& &\int\,\frac{d \Omega_6}{p^2+q^2-2\,p\,q\,\cos\theta}\,=\,
\frac{4 \,\pi^3}{3}\Biggl(\Theta(x-y)\biggl(\frac{3}{4 x}-
\frac{y}{4 x^2}\biggr)+\Theta(y-x)\biggl(\frac{3}{4 y}-
\frac{x}{4 y^2}\biggr)\Biggr)\,;\nn\\
& &\int\,d\Omega_6\,\log(p^2+q^2-2\,p\,q\,\cos\theta)\,=\,
\frac{\pi^3}{12}\Biggl(\Theta(x-y)\biggl(\frac{8 y}{x}-
\frac{y^2}{x^2}+12\,\log\,x\biggr)+\nn\\
& &+\,\Theta(y-x)\biggl(\frac{8 x}{y}-\frac{x^2}{y^2}+ 12\,\log\,y
\biggr)\Biggr)\,;
\label{angle}\\
& &\int\,d\Omega_6\,(p\,q)\,\log(p^2+q^2-2\,p\,q\,
\cos\theta)\,=\,
\frac{\pi^3}{18}\Biggl(\Theta(x-y)\biggl(\frac{3 y^2}{x}-6\,
y-\frac{3 y^3}{5 x^2}\biggr)+\nn\\
& &+\,\Theta(y-x)\biggl
(\frac{3 x^2}{y}-6\,x-\frac{3 x^3}{5 y^2}\biggr)\Biggr)\,;
\nn\\
& &x\,=\,p^2\,;\qquad y\,=\,q^2\,.\nn
\eeq
First of all let us calculate one-loop integral keeping terms of 
zero and the first orders in $ m^2$. We have for 
one such vertical diagram ($x = p^2$, 
where $p$ is the total momentum along the loop)
\be
-\,\imath\,\frac{G^2\, \pi^3}{2\,(2 \pi)^6}\,\Biggl(
\Lambda^2\,+\,\frac{1}{3}\,x\,\log\biggl(\frac{x}{\Lambda^2}
\biggr)\,+2\,m^2\,\log\biggl(\frac{x}{\Lambda^2}
\biggr)\,-\,c\,x\Biggr)\,;\label{loop}
\ee
where $\Lambda$ is the square of the cut-off being mentioned 
 and $c$ is a constant, depending on a behaviour of the form-factor.
\bigskip
\bigskip

\begin{picture}(160,85)
{\thicklines
\put(20,80.5){\line(-1,1){5}}
\put(20,80.5){\line(1,1){5}}
\put(20,80.5){\circle*{3}}
\put(20,80.5){\line(-1,-1){5}}
\put(20,80.5){\line(1,-1){5}}
\put(30,80){=}
\put(36,80){$G\,F(p)$}
{\thicklines
\put(80,80.5){\line(-1,1){5}}
\put(80,80.5){\line(1,1){5}}
\put(80,80.5){\circle*{1}}
\put(80,80.5){\line(-1,-1){5}}
\put(80,80.5){\line(1,-1){5}}}
\put(90,80){=}
\put(98,80){$G$}

{\thicklines
\put(5,50.5){\line(-1,1){5}}
\put(5,50.5){\line(1,1){5}}
\put(5,50.5){\circle*{3}}}
\put(5,50.5){\line(-1,-1){5}}
\put(5,50.5){\line(1,-1){5}}

\put(12.5,50){+}
{\thicklines
\put(22.5,50.5){\line(-1,1){5}} \put(32.5,50.5)
{\oval(20,10)[t]}
\put(42.5,50.5){\line(1,1){5}}\put(22.5,50.5)
{\circle*{1}}
\put(42.5,50.5){\circle*{3}}}
\put(22.5,50.5){\line(-1,-1){5}} \put(42.5,50.5)
{\line(1,-1){5}}
\put(22.5,50.5){\line(1,0){20}}
\put(53,50){+}
{\thicklines
\put(65,50.5){\line(-1,1){5}} \put(75,50.5)
{\oval(20,10)[t]}
\put(85,50.5){\line(1,1){5}}\put(65,50.5){\circle*{1}}
\put(85,50.5){\circle*{1}}}
\put(65,50.5){\line(1,-1){5}} \put(85,50.5)
{\line(-1,-1){5}}
\put(65,50.5){\line(1,0){20}}
\put(95.5,50){+}

{\thicklines
\put(115.5,60.5){\line(-1,1){5}}
\put(115.5,60.5){\line(1,1){5}}
\put(115.5,50.5){\oval(10,20)}
\put(115.5,60.5){\circle*{1}}
\put(115.5,40.5){\circle*{1}}}
\put(115.5,40.5){\line(-1,-1){5}}
\put(115.5,40.5){\line(1,-1){5}}
\put(130,50){+}
\put(0,10.5){+}
{\thicklines
\put(12.5,10.5){\line(-1,1){5}} \put(22.5,10.5)
{\oval(20,10)[t]}
\put(12.5,10.5)
{\circle*{1}}
\put(32.5,10.5){\circle*{1}}}
\put(12.5,10.5){\line(-1,-1){5}} 
\put(12.5,10.5){\line(1,0){20}}
{\thicklines
 \put(42.5,10.5)
{\oval(20,10)[t]}
\put(52.5,10.5){\line(1,1){5}}\put(32.5,10.5)
{\circle*{1}}
\put(52.5,10.5){\circle*{3}}}
 \put(52.5,10.5)
{\line(1,-1){5}}
\put(32.5,10.5){\line(1,0){20}}
\put(62.5,10.5){+}
{\thicklines
\put(100,10){\line(-2,1){30}}
\put(100,10){\line(-2,-1){30}}
\put(80,10){\oval(5,20)}
\put(80,20){\circle*{1}}
\put(80,0){\circle*{1}}
\put(100,10){\line(1,1){10}}
\put(100,10){\line(1,-1){10}}
\put(100,10){\circle*{3}}}}
\put(120,10){=}
\put(130,10){{\Large 0}}
\end{picture}
\bigskip
\bigskip

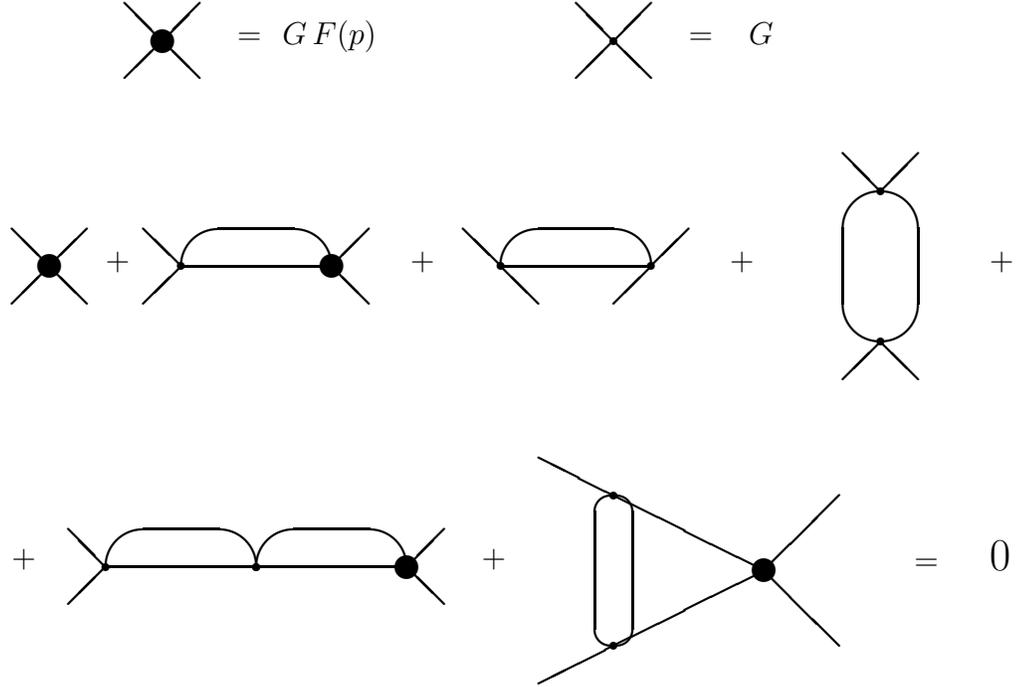
Fig. 1. The graphic representation of the linear compensation 
equation (18).\\

Let us consider the linear compensation equation, 
obtained in agreement with the formulated rules (see Fig. 1). 
The equation in this approximation has the following form
\beq
& &G\,F(p^2)\,=\,\frac{G^2}{2 (4 \pi)^3}\,\Biggl(3\,
\Lambda^2\,+\,\frac
{2}{3}\,p^2\,\log\biggl(\frac{p^2}{\Lambda^2}\biggr)\,+4\,m^2\,
\log\biggl
(\frac{p^2}{\Lambda^2}\biggr)\,-\,2\,c\,p^2\Biggr)\,\,-\nn\\
& &-\,\frac{G^3\,}{8\,(2 \pi)^{9}}\,
\int \Bigl( \frac{1}{3}(p-q)^2\,\log\frac{(p-q)^2}{\Lambda^2}\,
+2\,m^2\,\log \frac{(p-q)^2}{\Lambda^2}\,
-\,c\,(p-q)^2\Bigr)\,\times\label{inteq}\\
& &\times\,\frac{F(q^2)}{(q^2+m^2)^2}\,d^6 q\,-\,\frac{3\,G^3\,
\pi^3\,\Lambda^2}{2\,(2 \pi)
^{12}}\,\int \frac{F(q^2)}{(q^2+m^2)^2}\,d^6q\,.\nn
\eeq
Firstly let us note, that trivial solution $G\,=\,0$ is evidently 
possible. In view of looking for a non-trivial solution we 
cancel the equation by $G$.
Performing here angle integrations by using 
formulas~(\ref{angle}) we obtain the following 
one-dimensional integral equation 
\beq
& &F(x) = \frac{G}{2 (4\pi)^3} \Biggl(3 \Lambda^2 +
\frac{2}{3}\, x \, \log\biggl(\frac{x}{\Lambda^2}\biggr)+ 4 m^2\, 
\log\biggl(\frac{x}{\Lambda^2}\biggr) - 2 c x
\Biggr) - \nn\\
& &-\,\frac{3\,G^2 \Lambda^2}{4 (4 \pi)
^6} \int_0^\infty \frac{y^2\,F(y)}{(y+m^2)^2}\,dy\,-\,\frac{G^2}{18 
(4 \pi)^6} \Biggl(
-\,\frac{1}{20\,x^2} \int_0^x \frac{y^5\,F(y)}{(y+m^2)^2}\,dy\,+\nn\\
& &+\frac{3}{4 x} 
\int_0^x \frac{y^4\,F(y)}{(y+m^2)^2}\,dy\,+\,3\,\log\,x\int_0^x 
\frac{y^3\,F(y)}{(y+m^2)^2}\,dy\,+
\,3\,x\,\log\,x\,\int_0^x\frac{y^2\,F(y)}{(y+m^2)^2}\,dy\,+\nn\\
& &+\,3\,\int_x^\infty \frac{y^3\,\log\,y\,F(y)}
{(y+m^2)^2}\,dy\,+\,x\,\int_x^\infty \frac{(4 + 3 \log\,y)\,y^2\,F(y)}
{(y+m^2)^2}\,dy\,+\nn\\
& &+
\,4\,\int_0^x \frac{y^3\,
F(y)}{(y+m^2)^2}\,dy\,+\,\frac{3\,x^2}{4}\,\int_x^\infty\,\frac{y\,
F(y)}
{(y+m^2)^2}\,dy\,-
\,\frac{x^3}{20}\,\int_x^\infty\,\frac{F(y)}{(y+m^2)^2}\,dy
\Biggr)\,+\label{eq}\\
& &-\,\frac{G^2\,m^2}{12 (4 \pi)^6} \Biggl(
-\,\frac{1}{x^2} \int_0^x \frac{y^4\,F(y)}{(y+m^2)^2}\,dy +\frac{8}
{x} 
\int_0^x \frac{y^3\,F(y)}{(y+m^2)^2}\,dy + 12\, \log x\int_0^x 
\frac{y^2\,F(y)}{(y+m^2)^2}\,dy +\nn\\
& &+\,12\,\int_x^\infty \frac{y^2\,\log\,y\,F(y)}
{(y+m^2)^2}\,dy\,+\,8\,x\,\int_x^\infty \frac{y\,F(y)}
{(y+m^2)^2}\,dy\,
-\,x^2\,\int_x^\infty\,\frac{F(y)}{(y+m^2)^2}\,dy\,\Biggr)+\nn\\
& &+\,\frac{G^2}{6\,(4\pi)^6}\,\biggl(\log\,
\Lambda^2\,+\,3\,c\biggr)\Biggl(
\int_0^\infty\,\frac{y^3\,F(y)}{(y+m^2)^2}\,dy\,+\,x\,\int_0^\infty\,
\frac{y^2\,F(y)}{(y+m^2)^2}\,dy
\Biggr)\,+\nn\\
& &+\,\frac{G^2\,m^2}{(4\pi)^6}\,\log\,
\Lambda^2\,\int_0^\infty\,
\frac{y^2\,F(y)}{(y+m^2)^2}\,dy\,.\nn
\eeq

A method of solution of equations of~(\ref{eq}) type 
is developed in work~\cite{AF}. Equation~(\ref{eq}) is 
reduced to a differential one by sequential differentiations. 
Evident evaluation gives
\beq
& &\frac{d^4}{d x^4}\Biggl( x^2 \frac{d^4}{d x^4} \biggl( 
x^2 F(x) \biggr)\Biggr) = -\,\beta \Biggl(\frac{F(x)}{(x+m^2)^2}
+2 m^2 \biggl(x\,\frac{d^2}{dx^2}\frac{F(x)}{(x+m^2)^2} + 3 
\frac{d}{dx}\frac{F(x)}{(x+m^2)^2} \biggr)\Biggr);\nn\\
& &\quad \beta\,=\,\frac{2\,G^2}{(4\,\pi)^6}\,.\label{deq}
\eeq
One easily see, that Eq.~(\ref{deq}) can be rewritten in 
the form
\beq
& &\Biggl(\biggl(x\,\frac{d}{dx}\,+\,2\biggr)
\biggl(x\,\frac{d}{dx}\,+\,1\biggr)
\biggl(x\,\frac{d}{dx}\biggr)
\biggl(x\,\frac{d}{dx}\biggr)
\biggl(x\,\frac{d}{dx}\,-\,1\biggr)
\biggl(x\,\frac{d}{dx}\,-\,1\biggr)\,\times\nn\\
& &\times \biggl(x\,\frac{d}{dx} - 2\biggr)
\biggl(x \frac{d}{dx} - 3\biggr)\,+
\,\beta\,x^2
\Biggr)\,F(x)\,=\,2\,\beta\,m^2 x \biggl(F(x) + 
x \frac{d\,F}{dx} - x^2 \frac{d^2\,F}{dx^2} \biggr);\label{deqm}
\eeq
where two terms of expansion in $m^2$ are kept. 
From this form of the equation we immediately conclude, 
that for $x \to 0$ there are eight independent 
asymptotes, which coefficients we denote as follows
\beq
& &\frac{a_{-2}}{x^2};\quad \frac{a_{-1}}{x};\quad 
a_0 \,;\quad a_{0l}\,\log\,x;\quad a_1\,x;\nn\\ 
& &a_{1l}\,x\,\log\,x;\quad a_2\,x^2;\quad 
a_3\,x^3\,.\label{as0}
\eeq
Eight independent asymptotes at infinity are the following
\be
F_k(x)\,\simeq\,x^{-3/8}\,\exp\Biggl(4\,(\beta\,x^2)^
{1/8}\exp\Bigl(\frac{\imath \,\pi\,(2 k-1)}{8}\Bigr)\Biggr)\,;
\qquad k\,=\,1,\,2,\,...,\,8.\label{asinf}
\ee
Four of these asymptotes at infinity decrease exponentially
($k\,=\,3,\,4,\,5,\,6$), and the rest four ones do increase.

Equation~(\ref{deqm}) is equivalent to the initial integral 
equation under definite boundary conditions being fulfilled. 
First of all we can use only solutions, decreasing at 
infinity. To obtain conditions at zero we have to 
substitute expression
$$
F(x)\,=\,-\,\frac{x^2}{\beta}\,\frac{d^4}{d x^4}\Biggl
(\,x^2\,\frac{d^4}{d x^4}\,\biggl(\,
x^2\,F(x)\,\biggr)\Biggr)\,;
$$
in integrals of equation~(\ref{eq}) and perform sequential 
integrations by parts. The results are presented in the Appendix. 

Substituting expressions~(\ref{integ}) into 
equation~(\ref{eq}), we have
\beq
& &F(x)\,=\,F(x)\,-\,\frac{a_{-2}}{x^2}\,-\,\frac{a_{-1}}{x}
\,-\,a_{0l}\,\log\,x\,-\,a_{1l}x\,\log\,x\,+\nn\\
& &+\,\frac{G\,\pi^3}
{2 (2\pi)^6}\,\Biggl(3\,\Lambda^2 \biggl(1- \frac
{G\,I}{2 (4\,\pi)^3}\biggr)\,+
\,\frac
{2\,x}{3}\log\biggl(\frac{x}{\Lambda^2}\biggr)\,-\,2\,
c\,x\Biggr)\,+\,x\,\biggl(\log\,
\Lambda^2\,+\,3\,c\biggr)\,a_{1l}\,;\label{eqb}\\
& &I\,=\,\int_0^\infty\frac{y^2\,F(y)}{(y+m^2)^2}\,dy\,.\nn
\eeq
From here we obtain the following condition (independently 
on values of $\Lambda^2$ and $c$)
\beq
& &a_{-2}\,=\,0\,,\qquad a_{-1}\,=\,0\,,\quad 
a_{0l}\,=\,\frac{2\,G\,m^2}{(4 \pi)^3}\,,\nn\\
& &a_{1l}\,=\,\frac{G \pi^3}{3\, (2\pi)^6}\,
=\,\frac{\sqrt{2\,\beta}}{6}\,;\label{bound} \\
& &I\,=\,\frac{2 (4 \pi)^3}{G}\,=\,\frac{2\,\sqrt{2}}
{\sqrt{\beta}}\label{I}
\,.
\eeq
The first four conditions~(\ref{bound}) are boundary 
conditions for Eq.~(\ref{deqm}). A combination 
of four solutions decreasing at infinity with account of 
these boundary conditions gives the unique 
solution. It can be expressed in terms of well-known special 
functions for case $m^2\,=\,0$. Indeed, let us make the following 
substitution in Eq.~(\ref{deqm})
\be
z\,=\,\frac{\beta\,x^2}{2^8}\,;\label{sub}
\ee
which reduces the equation to the canonical form of 
Meijer equation~\cite{be} of the eighth order
\beq
& &\Biggl(\biggl(z\,\frac{d}{dz}\,+\,1\biggr)
\biggl(z\,\frac{d}{dz}\,+\,\frac{1}{2}\biggr)
\biggl(z\,\frac{d}{dz}\biggr)
\biggl(z\,\frac{d}{dz}\biggr)
\biggl(z\,\frac{d}{dz}\,-\,\frac{1}{2}\biggr)
\biggl(z\,\frac{d}{dz}\,-\,\frac{1}{2}\biggr)\,\times\nn\\
& &\times\,\biggl(z\,\frac{d}{dz}\,-\,1\biggr)
\biggl(z\,\frac{d}{dz}\,-\,\frac{3}{2}\biggr)\,+
\,z\,
\Biggr)\,F(z)\,=\,0\,.\label{canon}
\eeq
Conditions~(\ref{bound}) fix the solution. Firstly, four 
solutions, decreasing at infinity, always could be combined 
to set to zero three singular asymptotes at zero, i.e. 
to fulfill conditions $a_{-2} = a_{-1} = a_{0l} = 0$. Such 
property has the following Meijer function (see~\cite{be})
$$
C\cdot G_{08}^{50}(\,z\,|\,3/2,\,1,\,1/2,\,1/2,\,0,\,0,\,-1/2,
\,-1).
$$
The constant is defined by the coefficient afore 
$\sqrt{z}\log\,z$. For small $z$ this Meijer function 
is such~\cite{be}
\be
G_{08}^{50}(z\,|\,3/2,\,1,\,1/2,\,1/2,\,0,\,0,\,-1/2,\,-1)
\,=\,\pi\,+\,
\frac{16}{3}\,\sqrt{z}\,\log\,z\,+\,...\,.\label{expan}
\ee
Comparing the coefficient afore $\sqrt{z}\,\log\,z$ 
with~(\ref{bound}), we obtain
$$
C\,=\,\frac{\sqrt{2}}{4}\,.
$$
Performing integration (see~\cite{Mar3}), we have in 
accordance with definition of $I$~(\ref{eqb}) 
\be
I = \int_0^\infty F(y)\,dy = \frac{\sqrt{2}}{4} 
\int_0^\infty G_{08}^{50}(\beta y^2/2^8\,|\,3/2,1,1/2,
1/2,0,0,-1/2,-1)\,dy = \frac{2 \sqrt{2}}
{\sqrt{\beta}};
\label{Lambda}
\ee
that perfectly agrees with condition~(\ref{bound}).

Thus, solution
\be
F(x)\,=\,\frac{\sqrt{2}}{4}\,
G_{08}^{50}(\beta x^2/2^8\,|\,3/2,\,1,\,1/2,\,1/2,\,0,\,0,
\,-1/2,\,-1)\,;\label{sol}
\ee
fulfills all conditions~(\ref{bound}), and consequently 
the initial equation~(\ref{eq}), which is an approximate 
compensation equation. This solution is a nontrivial 
solution, which contains dimensional parameter $G$, and 
hence it leads to the initial scale symmetry breaking. 
Of course as we have noted before trivial solution $F(x)\,=\,0$ 
is also possible. Note, that the boundary conditions are not 
dependent on value of the form-factor at zero. Equality 
$F(0) = 1$ will serve as an additional condition in what 
follows.

Let us take into account terms proportional to $m^2$. We shall look 
for a correction to the solution of Eq.~(\ref{deqm}) in 
the following form
\be
F(x)\,=\,F_0(x)\,+\,\Delta\,F(x)\,.\label{Delta}
\ee
Substituting~(\ref{Delta}) into equation~(\ref{deqm}) we have 
the following equation in the first order in $m^2$
\beq
& &\Biggl(\biggl(x\,\frac{d}{dx}\,+\,2\biggr)
\biggl(x\,\frac{d}{dx}\,+\,1\biggr)
\biggl(x\,\frac{d}{dx}\biggr)
\biggl(x\,\frac{d}{dx}\biggr)
\biggl(x\,\frac{d}{dx}\,-\,1\biggr)
\biggl(x\,\frac{d}{dx}\,-\,1\biggr)\biggl(x\,\frac{d}{dx}\,-
\,2\biggr)\,\times\nn\\
& &\times\,
\biggl(x\,\frac{d}{dx}\,-\,3\biggr)\,+
\,\beta\,x^2\,
\Biggr)\,\Delta\,F(x)\,=\,2\,\beta\,m^2\,x\,\biggl(F_0(x)\,
+\,x\,\frac{d\,F_0}{dx}\,-\,x^2\,\frac{d^2\,F_0}{dx^2}\,\biggr).
\label{deqit}
\eeq

From equation~(\ref{deqit}) we can exactly define several terms of 
expansion of $\Delta\,F(x)$ for small $x$. Indeed let us consider the 
following expression
\beq
& &\bar \Delta\,F(x)\,=\,\frac{2\,m^2}{x}\,\biggl(F_0(x)\,
+\,x\,\frac{d\,F_0}{dx}\,-\,x^2\,\frac{d^2\,F_0}{dx^2}\,\biggr)\,=
\label{iter}\\
& &=\,2\,m^2\Biggl(\frac{\pi}{2 \sqrt{2}\,x} + 
\frac{2\,G}{3\,(4\pi)^3}
\biggl(\log\,(\sqrt{\beta}\,x)\,+4\,\gamma - \frac{23}{6}\biggr)\,
+\frac{\pi \sqrt{2}\, G^2}{96 \,(4\pi)^6}\,x\,\log\,x +O(x)\Biggr)\,;
\nn 
\eeq  
where $\gamma\,=\,0.577215665...\,$ is the Euler constant. 
Substituting 
expression~(\ref{iter}) into equation~(\ref{deqit}), we get 
convinced, 
that it fulfills the equation up to terms of $x^3$ order, because 
the differential operator in the left-hand side nullifies the 
terms presented in~(\ref{iter}) and subsequent terms up to the 
indicated order. We are interested just in the presented 
terms~(\ref{iter}) because they refer to the boundary conditions. 
Indeed expression~(\ref{iter}) contains terms $z^{-1/2},\,
\log z,\,z^{1/2}\,\log\,z$, which violate their boundary conditions. 
Hence we are to add to expression~(\ref{iter}) a combination of 
solutions of the homogeneous equation to force the boundary 
conditions to be fulfilled. Finally we obtain
\beq
& &\Delta\,F(x)\,=\,\bar \Delta\,F(x)\,-\,\frac{\pi^2\, Y}{8}\,
G^{50}_{08}
(\beta\,x^2/2^8\,|\,
3/2,\,1,\,1/2,\,1/2,\,-1/2,\,0,\,0,\,-1\,)\,-\nn\\
& &-\,\frac{2\,Y}{3}\,G^{50}_{08}(\beta\,x^2/2^8\,|\,
3/2,\,1,\,1/2,\,0,\,0,\,1/2,\,-1/2,\,-1\,)\,-\nn\\
& &-\,\pi\,Y\,\biggl(\gamma\,+\,\log\,2\,-\,\frac{43}{48}\biggr)\,
G^{50}_{08}(\beta\,x^2/2^8\,|\,
3/2,\,1,\,1/2,\,1/2,\,0,\,0,\,-1/2,\,-1\,)\,;\label{DF}\\
& &
\quad Y\,=\,\frac{G\,m^2}{2\,(4 \pi)^3}\,.\nn
\eeq
From this expression we extract the exact value for $F(0)$. While 
doing 
this one has to bear in mind, that the presence of a term  being 
proportional to $\log\,x$ at $x \to 0$ is a consequence of an 
expansion in $m^2$ at $x \gg m^2$. Looking back at the 
corresponding evaluations we see, that for $x \to 0$ one has 
to change $\log\,x$ for $\log 4\,m^2$. Now we have
\be
F(0)\,=\,\frac{\pi\,\sqrt{2}}{4}\,+\,Y\,\biggl( 4 \log Y+(16-\pi^2) 
\gamma 
+(14-\pi^2) \log 2-\frac{122}{9}+\pi^2\frac{42}{48}\biggr)\,.
\label{F00}
\ee

For $Y\,=\,0$ we obtain $F(0)\,=\,1.11072$. Condition $F(0)\,=\,1$ 
defines the value of $Y$, which is connected with mass 
(see~(\ref{DF}))
\be
Y\,=\,0.005789\,.\label{Y}
\ee
Thus the solution, which is found here, satisfies all the 
necessary conditions provided~(\ref{Y}) is valid. Emphasize, 
that~(\ref{Y}) defines the mass of the scalar field. Note, that 
the small value of~(\ref{Y}) thoroughly justifies the account 
of only the first term of the expansion in $m^2$. We reject the 
second solution of condition $F(0)=1$, which is of order of unity, 
due to to its inconsistence with the expansion of the solution in 
$m^2$. 

We have mentioned already, that generally speaking one has 
to consider a total chain of compensation equations 
including connected Green functions with six, eight, etc. 
legs. Note, that corresponding equations will contain 
inhomogeneous parts, expressed in terms of Green functions 
of lower order, and homogeneous parts, being proportional to 
the corresponding form-factor, e.g. $F_6$ with six legs. 
Assuming our result the connected four-leg Green function 
be zero, we come to the conclusion, that inhomogeneous 
part of equation for $F_6$ is zero, so trivial 
solution $F_6 = 0$ inevitably exists. The analogous 
considerations lead to conclusions on possibility of 
existence of trivial solutions of all higher Green functions. 
One may, of course, study possibilities of existence of 
nontrivial solutions as well. However, the purpose of the 
present work is to show that even though one nontrivial 
solution does exist, so we rely on following variant: 
nontrivial solution for four-leg connected Green function 
and trivial solutions for all higher connected  Green 
functions. The consideration of compensation equation 
for Green function with two legs, which defines mass of 
the scalar field will be performed particularly later on.

The next step of study should include non-linear equation with 
account 
of all possible diagrams. However this problem evidently do not admit 
analytic solution. Approximate estimate of non-linear corrections 
to the form-factor's value at zero will be obtained in what follows.
Maybe future studies will be connected with numerical methods. 
We are convinced, that the experience achieved in finding of the 
non-trivial solution will help in formulation and realization of 
numerical methods. Presumably result~(\ref{Y}), which means 
the existence of a solution only for definite relation between 
dimensional coupling constant and mass of scalar field, will 
be important.

\section{Bethe-Salpeter equation and zero excitation}

It is well-known, that a symmetry breaking is to be 
accompanied by an appearance of an excitation with 
zero mass~\cite{Bog2, Bog, Gold}. Let us consider this 
problem in the same approximation. While constructing 
an equation for a bound state one has to keep in mind, 
that here genuine interaction~(\ref{int}) acts, that one, 
which is referred to the interaction Lagrangian and 
remains, of course, not compensated.
Bethe-Salpeter equation for a massless bound state 
of two scalar fields in this case has the form
\beq
& &\Psi(x)\,=\,\frac{G\,\pi^3\,\Lambda^\prime }{2 (2\pi)^6}
\,-\,\frac{G^2\,\pi^6\,\Lambda\,\Lambda^\prime}{2\,(2 \pi)
^{12}}\,
+\,\frac{G^2\,\pi^6}{18\,(2 \pi)^{12}}\,\biggl(
-\,\frac{1}{20\,x^2}\,\int_0^x y^3\,\Psi(y)\,dy\,+\nn\\
& &+\,\frac{3}{4 x}\,
\int_0^x y^2\,\Psi(y)\,dy\,+\,3\,\log\,x\,\int_0^x y\,
\Psi(y)\,dy\,
+\,\nn\\
& &+\,3\,x\,\log\,x\,\int_0^x\,\Psi(y)\,dy\,+\,4\,\int_0^x y\,
\Psi(y)\,dy\,+\,3\,\int_x^\infty y\,\log\,y\,\Psi(y)\,dy\,
+\nn\\
& &+\,x\,\int_x^\infty (4 + 3 \log\,y)\,\Psi(y)\,dy\,+\,
\frac{3\,x^2}{4}\,\int_x^\infty\,\frac{\Psi(y)}{y}\,dy\,-\,
\frac{x^3}{20}\,\int_x^\infty\,\frac{\Psi(y)}{y^2}\,dy
\Biggr)\,;\label{eqBS}\\
& &\Lambda^\prime\,=\,\int_0^\infty \Psi(y)\,dy\,.\nn
\eeq
Comparing this equation~(\ref{eqBS}) with compensation 
equation~(\ref{eq}), we see the main difference in the sign 
afore the kernel of the integral equation. Remind once more, 
that the compensation equation is the condition of vanishing 
of the total expansion in $G$ in the modified free Lagrangian 
in expression~(\ref{phi}) and therefore terms of the first 
and of the third orders are situated in the same part of 
equation, e.g. in the left-handed one, whereas in the 
Bethe-Salpeter equation the corresponding terms are 
situated in different parts of equation. 
   
The sign afore the kernel is very important. This means, 
that in a differential equation sign afore $\beta$ changes 
as well
\beq
& &\Biggl(\biggl(x\,\frac{d}{dx}\,+\,2\biggr)
\biggl(x\,\frac{d}{dx}\,+\,1\biggr)
\biggl(x\,\frac{d}{dx}\biggr)
\biggl(x\,\frac{d}{dx}\biggr)
\biggl(x\,\frac{d}{dx}\,-\,1\biggr)
\biggl(x\,\frac{d}{dx}\,-\,1\biggr)\,\times\nn\\
& &\times\,\biggl(x\,\frac{d}{dx}\,-\,2\biggr)
\biggl(x\,\frac{d}{dx}\,-\,3\biggr)\,-
\,\beta\,x^2\,
\Biggr)\,\Psi(x)\,=\,0\,.\label{deqBS}
\eeq
One easily see, that due to absence of term being 
proportional to $x\,\log\,x$ in the inhomogeneous part 
boundary conditions are the following
\be
a_{-2}\,=\,a_{-1}\,=\,a_{0l}\,=\,a_{1l}\,=\,0\,.
\label{boundpsi}
\ee
The change of sign afore $\beta$ leads to 
changing of asymptotes at infinity
\be
\Psi_k(x)\,\simeq\,x^{-3/8}\,\exp\Biggl(4\,(\beta\,x^2)^
{1/8} \exp\Bigl(\frac{\imath \,\pi\, k}{4}\Bigr)\Biggr)\,;
\qquad k\,=\,1,\,2,\,...,\,8.\label{asinfpsi}
\ee
Now we have three decreasing asymptotes ($k = 3,\,4,\,5$), 
two oscillating ones with power decreasing ($k = 2,\,6$), 
and the remaining three are increasing. Using the first five 
solutions, which allow a definition of integrals at infinity, 
we fulfill four boundary conditions at zero~(\ref{boundpsi}). 
AS a result we obtain the following solution of 
equation~(\ref{eqBS})
\be
\Psi(x)\,=\,A\,G^{40}_{08}(\beta\,x^2/2^8\,|\,3/2,\,1,\,1/2,
\,0,\,1/2,\,0,\,-1/2,\,-1)\,;\label{solpsi}
\ee
where constant $A$ is defined by normalization condition of 
a Bethe-Salpeter wave function. Direct 
calculation~\cite{Mar3} leads to result $\Lambda^\prime\,
=\,0$, so the inhomogeneous part of equation~(\ref{eqBS}) 
vanishes. Thus we have shown, that the equation for a 
bound state with zero mass has a solution.

The solution being obtained proves the existence of zero mass 
excitation~\cite{Bog2, Bog, Gold} in the model. Of course definition 
of a Bethe-Salpeter equation itself is possible only provided a 
non-trivial solution of a compensation equation to exist and thus 
interaction~(\ref{int}) to act. The obligatory correspondence between 
a non-trivial solution of a compensation equation and an existence 
of a zero excitation thoroughly corresponds to Bogolubov 
quasi-averages approach~\cite{Bog}.

It is interesting to note, that with taking into account of 
three-fold interaction $g\,\phi^3$ in the kernel of 
equation~(\ref{eqBS}) 
the mass of the bound state becomes non-zero. One easily understands 
this, 
because interaction ~(\ref{eqBS}) itself leads to dimensional 
parameter 
$\Lambda_3$ being present and thus the scale invariance being already 
broken.

\section{Compensation equation for scalar field mass}

Let us look at interaction Lagrangian~(\ref{0i}). The mass term there 
is quite improper. To solve the problem one has to formulate 
a compensation equation for Green function with two scalar legs. Let 
us 
consider this equation taking into account solution~(\ref{Delta}) and 
three-fold interaction. The compensation equation means nullification 
of total contribution of interaction~(\ref{0i}) to the mass. In the 
first approximation the contribution of the four-fold interaction 
is described by the first order diagram "bubble" and that of the 
three-fold one is represented by simple one-loop diagram 
(see Fig. 2). 

\begin{picture}(160,30)
{\thicklines
\put(10,0){{\Large $m^2$}}
\put(25,0){=}
\put(40,0){\line(1,0){40}}}
\put(60,0){\circle*{3}}
\put(60,10){\oval(10,20)}

\put(90,0){+}
{\thicklines 
\put(120,0){\oval(20,10)[t]}
\put(130,0){\circle*{1}}
\put(110,0){\circle*{1}}
\put(100,0){\line(1,0){40}}}
\end{picture}
\bigskip
\bigskip

Fig. 2. Compensation equation for mass of the scalar field.\\ 

Putting  
momenta of the external legs to be zero, we have for "bubble" 
diagram just solution~(\ref{Delta}) in the vertex. As a result we 
obtain the following compensation equation for scalar mass
\beq
& &m^2\,=\,-\,\frac{G}{(2\,\pi)^6}\,\int\,\frac{F(q^2)\,d^6 q}{q^2 + 
m^2}\,-\,\frac{g^2}{(2 \pi)^6}\int\frac{d^6 q}{(q^2+m^2)^2}\,
=\nn\\
& &=\,-\,\frac{G}{2\,(4\,\pi)^3}\,
\int_0^\infty\,y\, dy\,(F_0(y)\,+\,\Delta F(y))\,+\,
\frac{G m^2}{2\,(4\,\pi)^3}\,
\int_0^\infty\,dy\,F_0(y)
\,-\nn\\
& &-\,\frac{g^2}{2\,(4 \pi)^3}\int_0^\infty\frac{y^2 \,dy}
{(y+m^2)^2}\,;\label{compm}
\eeq
Here in "bubble" diagram we perform an expansion in $m^2$ and take 
into 
account the zeroth and the first orders of the expansion. By direct 
evaluation 
with the aid of expressions~(\ref{sol}, \ref{iter}, \ref{DF}) we 
obtain 
that the zeroth order terms is zero and the first order term is equal 
to $3\,m^2$. The loop, which is described by the last term 
in~(\ref{compm}), 
quadratically diverges. Note, that in the initial theory~(\ref{init}) 
we introduce some cut-off $\Lambda_3$, which corresponds to 
a physical limitation of a region of applicability of the theory. 
As a result we have the following compensation 
equation for the mass provided $m \ll \Lambda_3$ 
\be
m^2\,=\,3\,m^2\,-\,\frac{g^2}{2\,(4\pi)^3}\,\Lambda_3^2.\label{hy}
\ee
Emphasize, that for the trivial solution $G = 0$ the first term in 
the 
right-hand side of equation~(\ref{hy}) is absent and we have a 
negative 
mass squared, i.e. a tachyon solution. For the non- trivial solution 
we 
have
\be
m^2\,=\,\frac{g^2}{4\,(4\,\pi)^3}\,\Lambda_3^2\,.\label{m}
\ee  
It is well-known, that a scalar tachyon leads to instability for 
small fields. Therefore the restoration of the normal sign of the 
mass 
squared, which is achieved provided the non-trivial solution is 
valid, 
corresponds to a transition to a more stable state.

So the value of the scalar mass is defined in terms of initial 
parameters of the theory $g$ and $\Lambda_3$. The value of 
parameter $Y$~(\ref{Y}) gives the relation of the mass and 
of the coupling constant $G$ of the four-fold interaction. Thus all 
the parameters entering into the non-trivial solution are defined in 
terms of the initial ones.

Note that the initial cut-off $\Lambda_3$ corresponds to some 
boundary energy, which provides real physical cut-off of the 
corresponding integrals. In the physical four-dimensional space-time 
it may be for example the Planck energy $1.22\cdot 10^{19}\, GeV$. 
Note also that in realistic models without elementary scalars 
(see e.g. \cite{Arb1}, \cite{SUSY}) quadratic divergences in mass of 
the elementary fields are absent. One should expect the expressions 
similar to~(\ref{hy}) also would lead to relations, which connect  
the theory parameters with a boundary energy (e.g. the Planck one), 
which enters into logarithmically divergent terms.

The final result for effective Lagrangian of the theory after 
the symmetry breaking occurs is the following
\be
L\,=\,\,\frac{1}{2}\,\frac{\partial \phi}{\partial x^\mu}\,
\frac{\partial \phi}{\partial x^\mu}\,-\,\frac{G}{4!}\,
\bar F(x_1,x_2,x_3,x_4)\,\phi(x_1)\phi(x_2)\phi(x_3)
\phi(x_4)\,;\label{phin}
\ee
where form-factor $F$ is the solution of the compensation equation.

\section{Estimate of non-linearity influence}

Till now our results were obtained in the framework of the linear 
approximation. The decrease of the form-factor at infinity indicates 
an applicability region of the approximation. It evidently is 
incorrect  
for large momenta variables because the effective coupling constant 
becomes too small in comparison to constant $G$, which was used to 
define the kernel of the integral equation. We can roughly take into 
account an influence of a non-linearity, using the following 
procedure.
Let equation~(\ref{deqm}) be valid for small $x$ (we put $m^2 = 0$). 
\be
\frac{d^4}{d x^4}\Biggl(\,x^2\,\frac{d^4}{d x^4}\,\biggl(\,
x^2\,F(x)\,\biggr)\Biggr)\,=\,-\,\beta\,\frac{F(x)}{x^2}\,;
\quad \beta\,=\,\frac{2\,G^2}{(4\,\pi)^6}\,.\label{deq0}
\ee
We use this equation with the corresponding boundary 
conditions~(\ref{bound}) for $x \le x_0$, whereas for $x \ge x_0$  
one has to take into account a non-linearity. Let us draw 
attention to the fact, that $\beta$ is proportional to $G^2$ i.e. 
it contains the form-factor squared. Therefore for  $x \ge x_0$ 
instead of~(\ref{deq0}) we use the following equation
\be
\frac{d^4}{d x^4}\Biggl(\,x^2\,\frac{d^4}{d x^4}\,\biggl(\,
x^2\,F(x)\,\biggr)\Biggr)\,=\,-\,\beta\,\frac{F^3(x)}{x^2}\,.
\label{deq2}
\ee
In this approximation we have correct behaviour of right-hand sides 
at small~(\ref{deq0}) and at very large~(\ref{deq2}) values of $x$.
In the intermediate region there is a tear in the rhs. at 
$x\,=\,x_0$. 
This means that the eighth derivative tears at this point. As we 
shall 
see soon the form-factor and its derivatives up to the fifth order 
have to be continuous. 

Let us introduce variable~$y\,=\,\sqrt{\beta}\,x$. One easily sees 
that for $y \to \infty$ equation~(\ref{deq2}) defines the following 
decreasing asymptotics
\be
F(y)\,\simeq\,\frac{b}{y^2}\,-\,\frac{6\,b^3}{5!\,7!\,y^4}\,+\,
\frac{12\,b^5}{7!\,7!\,8!\,y^6}\,+\;...\,;\label{asymn}
\ee
where $b$ is a constant. At the same time equation~(\ref{deq0}) 
with account of boundary conditions has the following solution 
in region $(0,\,y_0)$ 
\beq
& &F(y)\,=\,\frac{\sqrt{2}}{4}\,G^{50}_{08}(y^2/256\,|\,3/2,\,1,\,
1/2,\,1/2,\,0,\,0,\,-1/2,\,-1)\,+\nn\\
& &+\,C_1\,G^{30}_{08}(y^2/256\,|\,3/2,\,1,\,1/2,\,1/2,\,0,
\,0,\,-1/2,\,-1)\,+\nn\\
& &+\,C_2\,G^{30}_{08}(y^2/256\,|\,3/2,\,1,\,0,\,1/2,\,1/2,
\,0,\,-1/2,\,-1)\,+\label{lin}\\
& &+\,C_3\,G^{10}_{08}(y^2/256\,|\,3/2,\,1,\,1/2,\,1/2,\,0,
\,0,\,-1/2,\,-1)\,+\nn\\
& &+\,C_4\,G^{10}_{08}(y^2/256\,|\,1,\,3/2,\,1/2,\,1/2,\,0,
\,0,\,-1/2,\,-1)\,;\nn         
\eeq
where $C_i$ are constants. The appearance of the additional terms 
with 
these coefficients multiplied by Meijer functions increasing at 
infinity 
is due to the fact, that now the decrease at infinity is provided by 
asymptotics~(\ref{asymn}) and thus in region $(0,\,y_0)$ we have to 
use 
all solutions of equation~(\ref{deq0}), which fulfill the boundary 
conditions at zero.  The first line here is solution~(\ref{sol}), 
which 
was obtained earlier. Let us begin a sequential account of the new 
terms 
starting from the zero approximation, in which in region $(0,\,y_0)$ 
we 
have this old solution, i.e. all $C_i\,=\,0$. This solution is 
matched 
to solution~(\ref{asymn}) in point $y_0$. It will come clear, that 
in   
expression~(\ref{asymn}) an account of the first term is sufficient. 
Then from continuity of the function and of its first derivative we 
obtain the following set of equations
\beq
& &\frac{\sqrt{2}}{4}\,G^{50}_{08}(y^2_0/256\,|\,3/2,\,1,\,1/2,\,1/2,
\,0,
\,0,\,-1/2,\,-1)\,-\,\frac{b}{y_0^2}\,=\,0\,;\nn\\
& &\frac{\sqrt{2}}{4}\,G^{50}_{08}(y^2_0/256\,|\,3/2,\,1,\,1,\,1/2,
\,1/2,
\,0,\,-1/2,\,-1)\,-\,\frac{b}{y_0^2}\,=\,0\,;\label{set0}
\eeq
Solution of the set:
\be
y_0\,=\,8.4980\,;\qquad b\,=\,7.5055\,.\label{sol0}
\ee
The second term in asymptotics~(\ref{asymn}) at $y_0$ 
comprises $7.7\cdot 10^{-6}$ times the first one, that justifies 
the account of the first term only. The value of the form-factor 
at zero does not change $F(0)\,=\,1.1107$. 

Now let us take into account two additional terms in~(\ref{lin}) 
with coefficients $C _1$ and $C_2$, which for small $y$ give 
larger contribution than the remaining two terms. In this case we 
have to match values of the function and of its derivatives up 
to the third order. One obtains the set of four equations with 
aid of rules of differentiation of Meijer functions~\cite{be}. 
Its solution reads
\be
y_0\,=\,17.635\,;\quad b\,=\,9.410\,;\quad C_1\,=\,0.0166\,;\quad 
C_2\,=\,-0.0538\,.\label{sol2}
\ee 
The value of the form-factor at zero becomes the following
\be
F(0)\,=\,\frac{\pi\, \sqrt{2}}{4}\,+\,\frac{C_2}{\pi}\,=\,1.0936\,.
\label{F01}
\ee

Now let us take into account terms with coefficients $C_3,\,C_4$.
We consider them and deviations from solution~(\ref{sol2}) as well 
to be small. Then matching the function and its derivatives up to 
the fifth order, we obtain a set of six linear equations leading 
to the following solution
\beq
& &\Delta y_0\,=\,1.457,\qquad \Delta b\,=\,1.032,\qquad 
\Delta C_1\,=\,-\,0.0094,\nn\\
& &\Delta C_2\,=\,0.0223,\qquad C_3\,=\,-\,0.0249,\qquad\; C_4\,=
\,0.0136\,.
\label{sol4}
\eeq 
Substituting the last result into~(\ref{F01}), we have
\be
F(0)\,=\,1.1007\,.\label{F02}
\ee
The sequence of numbers 1.1107, 1.0936, 1.1007 for value $F(0)$ 
demonstrates stability of the result in respect to contribution of 
non-linear corrections

\section{Conclusion} 

Grounding on the results being obtained we express a hypothesis, 
that in the model under consideration a nontrivial solution does 
exist, which breaks the initial scale invariance and leads to a 
spontaneous appearance of effective interaction in 
Lagrangian~(\ref{phin}), acting in a restricted region of the 
momenta space in accordance with the value of parameter $G$.  
Effective form-factor $F(p)$ decreases exponentially with 
oscillations 
for $p^2\,\to\,\pm\,\infty$, i.e. both for space-like and time-like 
momenta. We confirm the existence of a zero mass excitation, which 
has to be 
present for an occurrence of spontaneous symmetry breaking. 

We start with the asymptotically free theory of a scalar field 
(in a six-dimensional space), and we obtain as a result 
the definite theory with interaction breaking scale symmetry. 
New dimensional parameters $G^{-1/2}$ and $m$ are proportional to 
parameter $\Lambda_3$, which defines the initial asymptotically 
free interaction. 
Let us emphasize once more, 
that the interaction being obtained is an effective one, 
that first of all is reflected in a presence of form-factor 
$F(p)$, which is just the solution of compensation equation 
of N.N. Bogolubov method. At momentum infinity the theory 
becomes asymptotically free again.

It is quite important, that the problem under consideration 
has a consistent solution only provided triple interaction 
$g\,\phi^3$ is acting. Really, albeit compensation 
equation~(\ref{inteq}) contains no 
contribution of this interaction, the non-zero scalar field mass 
appears only for $g\,\ne\,0$. If it is not the case the value of 
form-factor at zero $F(0)$ is not unity. In general one can not 
exclude a possibility of condition $F(0) = 1$ being fulfilled 
for $m = 0$. However the experience obtained in considering the 
present problem shows that this condition could be fulfilled 
only provided the model has very peculiar properties. As a matter 
of fact the problem under consideration is defined not by 
compensation equation~(\ref{inteq}) only, but by set of 
equations~(\ref{inteq}, \ref{compm}), which explicitly contains 
a contribution of triple interaction $g\,\phi^3$.   

It should be noted, that a possibility of a nontrivial solution 
strongly depends on the choice of the theory. This may be 
demonstrated by comparison of different signatures of 
the six-dimensional space-time. Namely if one instead of 
signature $1\,+\,5$ will choose signature $3\,+\,3$, then 
in definition~(\ref{eucl}) of a transition to Euclidean 
coordinates the sign afore $\imath\,d^6 p$ changes. As a result 
all signs change for one-loop integrals. For four-fold 
interaction we restore all previous results by simple 
substitution $G \to -\,G$. However the one-loop integral 
with two three-fold vertices inevitably changes sign and 
relation~(\ref{hy}) leads to tachyon mass. So we come to 
the conclusion, that for signature $3\,+\,3$ only the trivial 
solution $G\,=\,0$ is stable.

Of course, we base our conclusions only on exact solutions 
of approximate equations. However it is possible, that qualitative 
properties of solutions, which manifest themselves in the model 
problem, will be quite useful in study of problems of 
spontaneous symmetry breaking in more realistic cases, when 
there is no hope for analytic solution of corresponding equations 
and what is possible to apply are just numerical methods. 
Attractive qualitative results are the existence of relations between 
parameters of the problem and the natural appearance of small 
parameter $Y$~(\ref{Y}). The essential result is connected also with 
the conclusion on the stability of the non-trivial solution. The 
estimate of non-linearity contribu\-ti\-on, which does not lead to 
decisive change of properties of the solution, provides 
additi\-on\-al argument on behalf of the present approach.

The resulting theory is non-local and the question might 
arise, whether the general principles of unitarity and 
causality are here valid. The initial theory~(\ref{init}) 
quite corresponds to these principles. One should expect, 
that its solutions, non-trivial  ones as well, have also to 
fulfill these conditions. Therefore one can consider 
the present example as a step in direction of formulating of a 
consistent non-local theory. Basing on results of the present 
work we may assume, that such theory can be consistent not for 
an arbitrary form-factor but for the one, which follows from a 
non-trivial solution of an initially local theory.
 
Without any doubt
a possibility of spontaneous appearance of an effective 
interaction, containing a dimensional parameter, is of great 
interest for studies of problems beyond the standard theory. 
In particular, the phenomenon of a spontaneous appearance of 
an effective interaction, provided it to occur in a genuine 
physical theory, e.g. in the electroweak theory, might 
essentially promote our understanding of bases of the theory. 
Some considerations in connection with this aspects are 
described in work~\cite{Arb1, SUSY}. A subsequent results in this 
direction and their connection with possible deviations from 
predictions of the standard theory will be presented elsewhere.  

The work is supported in part by grant RFBR 01-02-16209 
and by grant "Universities of Russia"$\,$ UR.02.03.002.

\appendix
\section{ Appendix }
Here formulas of integration by parts of expressions entering 
in equation~(\ref{eq}) are presented
\beq
& &\beta \int_0^x\frac{y^2\,F(y)}{(y+m^2)^2}\,dy\,=\,-x^2\,\frac{d^3}
{dx^3}\Bigl(
x^2\,\frac{d^4}{dx^4}\Bigl(x^2\,F(x)\Bigr)\Bigr)\,+\nn\\
& &+\,2 x\,\frac{d^2}{dx^2}\Bigl(
x^2\,\frac{d^4}{dx^4}\Bigl(x^2\,F(x)\Bigr)\Bigr)\,
-\,2\,\frac
{d}{dx}\Bigl(
x^2\,\frac{d^4}{dx^4}\Bigl(x^2\,F(x)\Bigr)\Bigr)\,+\,
12\,a_{1l}\,;\nn\\
& &\beta\int_0^x\frac{y^3\,F(y)}{(y+m^2)^2}\,dy\,=\,-x^3 \frac{d^3}
{dx^3}
\Bigl(
x^2 \frac{d^4}{dx^4}\Bigl(x^2 F(x)\Bigr)\Bigr)\,+\,3 x^2 
\frac
{d^2}{dx^2}\Bigl(
x^2 \frac{d^4}{dx^4}\Bigl(x^2 F(x)\Bigr)\Bigr)\,-\nn\\
& &-\,6 x\,\frac
{d}{dx}\Bigl(
x^2\,\frac{d^4}{dx^4}\Bigl(x^2\,F(x)\Bigr)\Bigr)\,+\,
6\,x^2\,\frac{d^4}{dx^4}\Bigl(x^2\,F(x)\Bigr)\,-\,
12\,a_{0l}\,;\nn\\
& &\beta \int_0^x\frac{y^4\,F(y)}{(y+m^2)^2}\,dy\,=\,-x^4 \frac{d^3}
{dx^3}
\Bigl(
x^2 \frac{d^4}{dx^4}\Bigl(x^2 F(x)\Bigr)\Bigr)\,+\,4 x^3 
\frac{d^2}{dx^2}\Bigl(
x^2 \frac{d^4}{dx^4}\Bigl(x^2 F(x)\Bigr)\Bigr)\,-\nn\\
& &-\,12 x^2\,\frac
{d}{dx}\Bigl(
x^2\,\frac{d^4}{dx^4}\Bigl(x^2\,F(x)\Bigr)\Bigr)\,+\,
24\,x^3\,\frac{d^4}{dx^4}\Bigl(x^2\,F(x)\Bigr)\,-\,24\,x^2\,
\frac{d^3}{dx^3}\Bigl(x^2\,F(x)\Bigr)\,+\nn\\
& &+\,
48\,x\,\frac{d^2}{dx^2}\Bigl(x^2\,F(x)\Bigr)\,-\,48\,
\frac{d}{dx}\Bigl(x^2\,F(x)\Bigr)\,+\,48\,a_{-1}\,;\nn\\
& &\beta \int_0^x\frac{y^5\,F(y)}{(y+m^2)^2}\,dy\,=\,-x^5 \frac{d^3}
{dx^3}
\Bigl(
x^2 \frac{d^4}{dx^4}\Bigl(x^2 F(x)\Bigr)\Bigr)\,+\,5 x^4\,
\frac{d^2}{dx^2}\Bigl(
x^2 \frac{d^4}{dx^4}\Bigl(x^2 F(x)\Bigr)\Bigr)\,-\nn\\
& &-\,20 x^3\,\frac
{d}{dx}\Bigl(
x^2\,\frac{d^4}{dx^4}\Bigl(x^2\,F(x)\Bigr)\Bigr)\,+\,
60\,x^4\,\frac{d^4}{dx^4}\Bigl(x^2\,F(x)\Bigr)\,-\,120\,x^3\,
\frac{d^3}{dx^3}\Bigl(x^2\,F(x)\Bigr)\,+\nn\\
& &+\,
360\,x^2\,\frac{d^2}{dx^2}\Bigl(x^2\,F(x)\Bigr)\,-\,720\,x\,
\frac{d}{dx}\Bigl(x^2\,F(x)\Bigr)\,+\,720\,x^2\,F(x)\,
-\,720\,a_{-2}\,;\nn\\
& &\beta \int_x^\infty\,\frac{F(y)}{(y+m^2)^2}\,dy\,=\,
\frac{d^3}{dx^3}
\Bigl(
x^2\,\frac{d^4}{dx^4}\Bigl(x^2\,F(x)\Bigr)\Bigr)\,;
\label{integ}\\
& &\beta \int_x^\infty\,\frac{y\,F(y)}{(y+m^2)^2}\,=\,x\,\frac{d^3}
{dx^3}
\Bigl(
x^2\,\frac{d^4}{dx^4}\Bigl(x^2\,F(x)\Bigr)\Bigr)\,-\,
\frac
{d^2}{dx^2}\Bigl(
x^2\,\frac{d^4}{dx^4}\Bigl(x^2\,F(x)\Bigr)\Bigr)\,;\nn\\
& &\beta\,\int_x^\infty\frac{y^2\,\log\,y\,F(y)}{(y+m^2)^2}\,dy\,=
\,x^2\,\log\,x\,
\frac{d^3}{dx^3}\Bigl(x^2 \frac{d^4}{dx^4}\Bigl(x^2 F(x)
\Bigr)\Bigr)\,-\,(2 x\log\,x+x) \times\nn\\
& &\times\,\frac{d^2}{dx^2}\Bigl(
x^2\,\frac{d^4}{dx^4}\Bigl(x^2\,F(x)\Bigr)\Bigr)\,+
\,(2\log\,x+3)\,\frac
{d}{dx}\Bigl(
x^2\,\frac{d^4}{dx^4}\Bigl(x^2\,F(x)\Bigr)\Bigr)\,-\nn\\
& &-\,2\,x\,
\frac{d^4}{dx^4}\Bigl(x^2\,F(x)\Bigr)\,+
\,2\,\frac{d^3}{dx^3}\Bigl(x^2\,F(x)\Bigr)\,;\nn\\
& &\beta \int_x^\infty\frac{y^2\,F(y)}{(y+m^2)^2}\,dy\,=\,x^2 
\frac{d^3}{dx^3}\Bigl(
x^2 \frac{d^4}{dx^4}\Bigl(x^2 F(x)\Bigr)\Bigr)\,-\,2 x 
\frac
{d^2}{dx^2}\Bigl(
x^2 \frac{d^4}{dx^4}\Bigl(x^2 F(x)\Bigr)\Bigr)\,+\nn\\
& &+\,2\,\frac
{d}{dx}\Bigl(
x^2\,\frac{d^4}{dx^4}\Bigl(x^2\,F(x)\Bigr)\Bigr)\,;\nn\\
& &\beta \int_x^\infty\frac{y^3\,\log\,y\,F(y)}{(y+m^2)^2}\,dy\,=
\,x^3\,\log\,x\,\frac{d^3}{dx^3}\Bigl(x^2 \frac{d^4}{dx^4}\Bigl
(x^2 F(x)\Bigr)\Bigr)\,-(3 x^2\log\,x+x^2) \times\nn\\
& &\times\,\frac{d^2}{dx^2}\Bigl(
x^2\,\frac{d^4}{dx^4}\Bigl(x^2\,F(x)\Bigr)\Bigr)\,+
\,(6\,x\log\,x+5 x)\,\frac
{d}{dx}\Bigl(
x^2\,\frac{d^4}{dx^4}\Bigl(x^2\,F(x)\Bigr)\Bigr)\,-\nn\\
& &-\,(6\log\,x+11)\,x^2\,
\frac{d^4}{dx^4}\Bigl(x^2\,F(x)\Bigr)\,+
\,6\,x\,\frac{d^3}{dx^3}\Bigl(x^2\,F(x)\Bigr)\,-\,
6\,\frac{d^2}{dx^2}\Bigl(x^2\,F(x)\Bigr)\,.\nn
\eeq

For equation~(\ref{eqBS}) one has to change sign afore
$\beta$ and change $F(x)$ for $\Psi(x)$. 
\end{document}